\documentclass[prl,nofootinbib, twocolumn,showpacs,aps,preprintnumbers,amsmath]{revtex4}
\usepackage{color}
\usepackage{amsmath,amsfonts,amssymb,amsthm,amstext,amscd,eucal,srcltx}
\usepackage{epsfig,graphicx,bm}
\usepackage{epstopdf, epsf}
\usepackage{dcolumn}% Align table columns on decimal point
\newcommand{\be}{\begin{equation}}
\newcommand{\ee}{\end{equation}}

\newcommand{\bse}{\begin{subequations}}
\newcommand{\ese}{\end{subequations}}
\newcommand{\bea}{\begin{eqnarray}}
\newcommand{\eea}{\end{eqnarray}}
\newcommand{\ba}{\begin{array}}
\newcommand{\ea}{\end{array}}
\newcommand{\bc}{\begin{center}}
\newcommand{\ec}{\end{center}}

\def\ads{AdS$_3$ }
\def\cft2{CFT$_2$ }

\def\tr{\tilde{r}}
\def\trho{\tilde{\rho}}
\def\tPhi{\tilde{\Phi}}
\def\z2{$\mathbb{Z}_2$ }

\makeatother

\begin{document}
\preprint{IPM/P-2010/009  \cr
          \eprint{arXiv:1003.4089} }
\vspace*{3mm}
\title{\emph{O-BTZ}:
Orientifolded BTZ Black Hole}
%\author{F. Loran$^1$ M. M. Sheikh-Jabbari$^2$}
%\affiliation {$^1$ Department of Physics, Isfahan University of
%Technology, Isfahan 84156-83111, Iran,\\
%$^2$School of Physics, Institute for research in fundamental
%sciences (IPM), P.O.Box 19395-5531, Tehran, Iran }
\author{F. Loran}
\affiliation {Department of Physics, Isfahan University of
Technology, Isfahan 84156-83111, Iran } \email{loran@cc.iut.ac.ir}
%jabbari@theory.ipm.ac.ir}
\author{M. M. Sheikh-Jabbari} \affiliation{$^2$School of Physics,
Institute for research in fundamental sciences (IPM), P.O.Box
19395-5531, Tehran, Iran} \email{jabbari@theory.ipm.ac.ir}

\begin{abstract}
Banados-Teitelboim-Zanelli (BTZ) black holes are constructed by
orbifolding  \ads geometry by boost transformations of its $O(2,2)$
isometry group. Here we construct a new class of solutions to \ads
Einstein gravity, orientifolded BTZ or \emph{O-BTZ} for short, which
in general, besides the usual BTZ orbifolding, involve orbifolding
(orientifolding) by  a \z2 part of $O(2,2)$ isometry group. This \z2
is chosen such that it changes the orientation on \ads while keeping
the orientation on its $2D$ conformal boundary.  O-BTZ solutions
exhaust all un-oriented \ads black hole solutions, as BTZ black
holes constitute all oriented \ads black holes.
O-BTZ, similarly to BTZ black holes, are stationary, axisymmetric
asymptotically \ads geometries with two
asymptotic charges, mass and angular momentum.

\end{abstract}
\date{\today}
\maketitle

%%%%%%%%%%%%%%%%%%%%%%%%%%%%%%%%%%%%%%%%%%%%%%%%%%%%%%%%%%%%%%%%%%%%%%%%%%%%%%%%%%%%%%%%%%%%%%%%%%%%%%%%%%%%%%%%%
\bc{\textit{\textbf{Introduction and Summary of results}}}\ec

Although does not have propagating gravitons, AdS$_3$ Einstein
gravity  has nontrivial black hole solutions, the BTZ black holes
\cite{BTZ1,BTZ2}, and hence provides a simple but at the same time
rich arena for addressing  questions about black holes and quantum
gravity in general. Since in three dimensions the number of
independent components of the Ricci tensor is the same as that of
Riemann curvature, all of the solutions to the equations of motion
for pure AdS$_3$ Einstein gravity are locally AdS$_3$ and are
obtained as quotients of {global} AdS$_3$ by (a subgroup of) its
$SO(2,2)$ isometry \cite{BTZ2}. For a generic BTZ black hole
solution, however, the quotient leads to closed time-like curves
(CTC's) in some regions of the space, which happen to fall behind
the inner horizon of the black hole geometry. In order to remove the
usual problems with the CTC's, those regions in the BTZ geometry are
cut out of the global AdS$_3$. As was discussed in \cite{BTZ2} and
we review below, the excised region includes a part of the causal
boundary of  global AdS$_3$ and renders the BTZ black holes as
geodesically incomplete.

Besides the geodesic incompleteness,  the BTZ solutions  also suffer
from a quantum instability. To see the instability one may introduce
a free scalar theory in the classical BTZ background geometry. It
has been shown that \cite{Steif}, see also \cite{ortiz-lifschytz},
expectation value of components of the energy momentum tensor of
quantum fluctuations of this field blows up at the inner horizon,
signaling an instability at the inner horizon, which is also a
Cauchy horizon in the (non-extremal) rotating BTZ black hole. This
problem seems to be a generic property of BTZ black holes and
independent of the details of the quantum field theory in question.
Noting that this instability is originating from inside the inner
horizon region \cite{Steif}, if one can cut this region in a
consistent manner, the problem with the instability, as well as the
geodesic incompleteness issue, might be resolved. However, so far it
is not clear how the latter should be performed.

It has been argued that  the AdS$_3$/CFT$_2$ provides us with the
tools to probe the region inside the horizon \cite{probing-inside}.
Computations with the CFT$_2$ again reveals the same instability at
the inner horizon of the BTZ black hole \cite{Inner-instability}.
The fact that CFT$_2$ correlators blow up at the inner (Cauchy)
horizon or at the (orbifold) singularity of the static BTZ has been
interpreted as impossibility of probing  the region beyond the
Cauchy horizon and hence a manifestation of the cosmic censorship
conjecture \cite{Inner-instability}.

On the other hand, in the case of pure Lorentzian AdS$_3$ Einstein
gravity, which is the case of our interest in this paper, we
\emph{may} have the advantage of a dual $2D$ conformal field theory
(CFT$_2$) description \cite{PureAdS3-CFT2}. This dual CFT$_2$, if it
exists, resides on the conformal (causal) boundary of AdS$_3$, which
is a $1+1 D $ flat space. The states of this \cft2 are labeled by
representations of a Virasoro algebra with equal left and right
central charges $c_L=c_R=c$. In the seminal work
\cite{Brown-Henneaux} Brown and Henneaux  showed that the Virasoro
algebra corresponding to the (proposed) dual CFT$_2$ is directly
related to a subset of $3D$ diffeomorphisms respecting certain
boundary
conditions, and %
\be\label{BH-central-charge} c=\frac{3\ell}{2G},
\ee%
where $\ell$ is  AdS$_3$ radius and $G$ is the $3D$ Newton constant.

The Brown-Henneaux boundary conditions \cite{Brown-Henneaux} allow
for diffemorphisms which change the orientation on the \ads while
preserving the orientation on the conformal boundary as well as
diffeomorphisms which preserve \ads orientation but change the
orientation on the boundary. The latter set of diffeomorphisms are
fixed by the choice of sign for energy and angular momentum in the
dual 2$D$ CFT and hence could be discarded. The former, however, may
be associated with a well-defined operator in the dual CFT. In this
Letter we will focus on such diffeomorphisms.

Motivated by the puzzles (features) of the usual BTZ solutions
discussed above, we construct the new class of stationary and
axisymmetric ``\emph{O-BTZ}'' solutions the line element of which
are the same as BTZ everywhere, with the same  mass and angular
momentum. O-BTZ black holes are obtained by orbifolding or
``orientifolding'' the orientation changing $\mathbb{Z}_2$ part of
the $O(2,2)$ isometry of AdS$_3$. This \z2 preserves the orientation
on the 2$D$ boundary of the AdS$_3$ and has a fixed locus which is a
{space-like cylinder} located in the region between the two horizons
of BTZ, the ``Space-like Orientifold Cylinder'' or SOC for short. As
such our O-BTZ black holes completes the results of \cite{BTZ1}:
\emph{All} possible black hole solutions to \ads Einstein gravity
are either BTZ or O-BTZ.

{}From the usual BTZ geometry viewpoint which serves as the
\emph{covering space} for the O-BTZ, hence, this \z2 exchanges the
region outside the outer horizon and the region inside the inner
horizon. Technically, the ``orientifold'' \z2 projection in the
covering space may be implemented by  cutting a BTZ geometry exactly
at the middle of the region between its two horizons (where the
fixed locus of the orientifold is located) and gluing another copy
of the same geometry to it. The O-BTZ geometry is hence the part of
BTZ geometry confined between the two orientifold fixed locus which
should be viewed as the surfaces at the end of the geometry. As
mentioned the O-BTZ everywhere except at the fixed locus of the
orientifold projection has the same metric as an ordinary BTZ. O-BTZ
geometry in the covering space may be viewed as a solution to \ads
Einstein gravity with a $\delta$-function source at the fixed locus
of the orientifold projection, the SOC.

In this way we obtain a solution which, despite of having the metric
of a rotating BTZ (outside its horizon), does not have an inner
horizon region. Therefore,  these solutions  do not have the CTC
issue (unlike the standard BTZ geometry). \footnote{The idea of
cutting the region with CTC's and gluing another part to the
geometry for removing the CTC problem has been   discussed, e.g. see
\cite{CTC-godel}. Note, however, that our construction, although
techincally and in the covering space of the orientifolding seems
similar, has the crucial difference that the geometry on the two
sides of the ``fixed locus'' are related by orinetifolding and have
exactly the same line element.} The analysis of \cite{Steif} can be
repeated for our geometry leading to the result that energy momentum
tensor for the fluctuations of any given field on the O-BTZ
backgrounds remain finite everywhere. The rest of this Letter is
organized as follows. We first  revisit and review some facts about
BTZ black holes. We then introduce the {O-BTZ} solutions and study
their causal structure and some other of their features. We end with
discussions and outlook.

%*******************************************************************************************************************************************
\vspace*{-2mm} \bc{\textit{\textbf{BTZ black holes revisited}}}\ec
\vspace*{-2mm}

\ads space is a hyperboloid embedded in
$R^{2+2}$:%
\be\label{AdS-embedding}%
 -T_1^2-T_2^2+X_1^2+X_2^2=-\ell^2\
. \ee%
A suitable coordinate for \ads is \cite{Steif}
\be\label{Steif-coordinates} ds^2=-({\tr^2}-{\ell^2})
dt^2+\ell^2 \frac{d\tr^2}{{\tr^2}-{\ell^2}}+\tr^2d\phi^2%
\ee%
where $t,\phi\in (-\infty,+\infty)$. For global \ads this coordinate
system should be extended past $\tr^2\geq 0$. This may be achieved
by replacing $\tr$ with $\trho$, $ \trho^2=\ell^2-\tr^2,$  where
$\tr^2$ becomes negative. In the $\tr^2\geq \ell^2$ region (region
I) $t$ is time-like and $\phi$ is space-like. In the region II,
where $\tr$ and $\trho$ coordinate systems overlap and $0 < \tr^2<
\ell^2$,  $t$ and $\phi$ are both space-like. In region III, where
$\trho^2>\ell^2$, $t$ coordinate becomes space-like while $\phi$ is
time-like. Relaxing $\tr^2>0$ condition, $\tr\leftrightarrow \trho,\
t\leftrightarrow \phi$ coordinate transformation is a $\mathbb{Z}_2$
diffeomorphism which exchanges regions I and III and maps region II
to itself. In the embedding $R^{2+2}$ space this $\mathbb{Z}_2$ may
be realized as
$%\be\label{Z2-transformation}%
X_1\longleftrightarrow X_2,\quad  T_1\longleftrightarrow T_2.
$
 This \z2 changes the orientation on the boundary while preserving
the \ads orientation. Later we will introduce another
$\mathbb{Z}_2$, the orientifold $\mathbb{Z}_2$, which changes the
orientation on AdS$_3$ while keeping the orientation on the 2$D$
boundary.

BTZ black holes \cite{BTZ1,BTZ2} constitute all  classical solutions
to the pure AdS$_3$ Einstein gravity (modulo the self-dual orbifold
\cite{self-dual-orbifold}) \emph{with a fixed \ads orientation} and
are obtained by orbifolding the original AdS$_3$ by the boosts of
its $SO(2,2)$ isometry:
\be\label{BTZ-identifications}%
\begin{split}
&T_1\pm X_1\equiv e^{\pm \frac{2\pi r_+}{\ell}}(T_1\pm X_1)\ , \cr
&T_2\pm X_2\equiv e^{\pm \frac{2\pi r_-}{\ell}}(T_2\pm X_2)\ .
\end{split}
\ee %
Without loss of generality we  assume $r_+> r_-\geq 0$. (For $r_+=r_-$ \eqref{BTZ-identifications} does not lead to a black hole \cite{BTZ2}.) For
$r_-=0$, the static BTZ black hole, the above orbifolding has a
fixed line at $T_1=X_1=0,\ T_2^2-X_2^2=\ell^2$ while for generic
$r_-\neq 0$ case the orbifolding is freely acting on \ads and
we have a smooth geometry. In the coordinate system
\eqref{Steif-coordinates} the BTZ identification
\eqref{BTZ-identifications} is written as%
\be\label{t-phi-identification}%
(t,\tr,\phi)\sim (t-2\pi r_-/\ell, \tr, \phi+2\pi r_+/\ell).%
\ee%

In the
 BTZ coordinates the metric takes the form%
\be\label{BTZ-coordinates}%
ds^2=\rho^2
d\tau^2+\frac{r^2dr^2}{16G^2J^2-\frac{r^2\rho^2}{\ell^2}}+r^2d\varphi^2 -8G\ell J
d\tau d\varphi\ ,
\ee%
where $\varphi\in [0,2\pi]$,%
\be%
\rho^2=8GM\ell^2 -r^2\ , \ %
\ee%
and $M, \ J$ are (ADM) mass and angular momentum
\be\label{AdS-MandJ}%
M=\frac{r_+^2+r_-^2}{8\ell^2 G},\quad J=\frac{r_+r_-}{4G\ell}\ .
\ee%
The  BTZ coordinate system \eqref{BTZ-coordinates} has the advantage
that the identification \eqref{BTZ-identifications} which constructs
the BTZ solution is performed only on the  $\varphi$ coordinate,
$\varphi\equiv \varphi+2\pi$.
%%%%%%%%%%%%%%%%%%%%%%%%%%%%%%%%%%%%%%%%%%%%%%%
\begin{figure}[t]
\includegraphics[scale=.95]{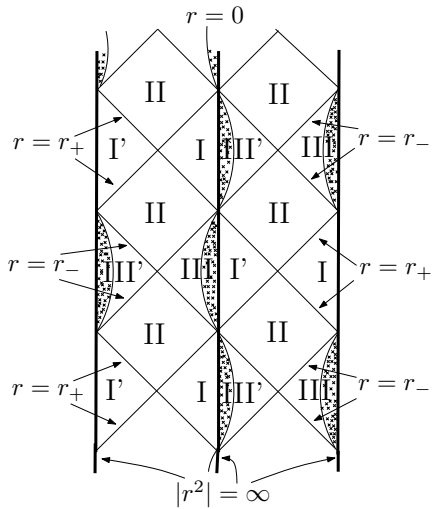}%
\caption{\label{BTZ} \textit{Two Penrose diagrams of a generic BTZ
black hole drawn side-by-side.} Regions I, I'  are  bounded between
the boundary at $r^2=+\infty$ and the outer horizon at $r=r_+$.
Region II is the region between the two horizons and Regions III,
III' are bounded between the inner horizon and boundary at
$\rho^2=\infty$. In the BTZ geometry the hatched area, corresponding
to $r^2<0$ is cut from the BTZ geometry. As depicted, this region
contains a part of the causal boundary of original AdS$_3$
\cite{BTZ2}. In the figure the $\varphi$ direction has been
suppressed and the $|r^2|=\infty$ lines correspond to
$\varphi=0,\pi, 2\pi$. To convey the idea that $r$ coordinate can be
extended past $r^2<0$ we have drawn two Penrose diagrams
side-by-side.}\vspace*{-4mm}
\end{figure}
%%%%%%%%%%%%%%%%%%%%%%%%%%%%%%%%%%%%%%%%%%%%%%

{}From identification \eqref{t-phi-identification} it is clear that
there are no closed time-like curves (CTC's) in the region where $t$
is parameterizing time. This is, however,  not sufficient and with
the above identification in the $r^2<0$ ($\rho^2>8GM\ell^2$) region,
which is inside the inner horizon of the BTZ black hole, geometry
develops CTC's. To remove inconsistencies arising from CTC's one is
forced to cut the geometry at $r=0$ \cite{BTZ2}. Penrose diagram of
the BTZ geometry is depicted in Fig.\ref{BTZ}.
%%%%%%%%%%%%%%%%%%%%%%%%%%%%%%%%%%%%%%%%%%%%%%%
\begin{figure}[t]
\includegraphics[scale=.95]{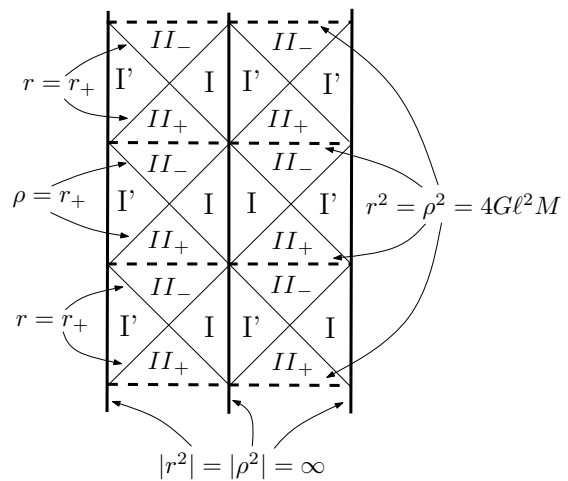}
\caption{\label{O-BTZ} \textit{Two Penrose diagrams of an O-BTZ
black hole in the ``orinetifold covering space'', drawn
side-by-side.}  $|r^2|=\infty$ line is the causal boundary which is a cylinder.ý
Fixed locus of the orientifold projection, the ``Space-like
Orientifold Cylinder'' (SOC),  is indicated by the dashed line
at $r^2=\rho^2=4G\ell^2 M$. As obtained by the \z2 projection, the regions above
and below the horizontal dashed line have the same metric.ý
The O-BTZ geometry is the part which is limited between two successive dashed lines. As it is seen the O-BTZ
geometry, is geodesically complete and does not have the inner
horizon or the region inside it. }\vspace*{-2mm}
\end{figure}
%%%%%%%%%%%%%%%%%%%%%%%%%%%%%%%%%%%%%%%%%%%%%%
\vspace*{-1mm}\bc{\textit{\textbf{ Orientifoleded-BTZ (O-BTZ)
solutions}}}\ec \vspace*{-1mm}

As mentioned, the only possible classical solutions to \ads Einstein
gravity should necessarily be locally \ads and  they are hence all
classified by orbifolding \ads by its isometries; which if we also
demand preserving the orientation on AdS$_3$, that is orbifolding
with a subgroup of $SO(2,2)$. These are BTZ solutions \cite{BTZ1,
BTZ2} or the self-dual \ads orbifold \cite{self-dual-orbifold}.

The only remaining possibility which we will study here is then to
orbifold (orientifold) \ads by the orientation changing \z2 which is
a part of $O(2,2)$ but not of $SO(2,2)$. Moreover, we would like
this orientifolding to also commute with that of BTZ
\eqref{BTZ-identifications}. Noting \eqref{t-phi-identification},
this is \emph{only} possible  if the orientifolding is acting on the
$\tilde{r}$ coordinate and not $t$ and $\phi$. Explicitly, that is
possible if the orientifold \z2 is $\tilde r^2\leftrightarrow \tilde
\rho^2$. This  \z2 does not have a simple (linear) realization on
$X_i$ and $T_i$ coordinates. In the BTZ coordinate system
\eqref{BTZ-coordinates} this
orientation changing  $\mathbb{Z}_2$ is hence%
\be\label{Orientation-Z2}%
(\tau, r^2, \varphi)\ \longleftrightarrow\ (\tau,\rho^2,\varphi) \ .
\ee%
(Note that $\varphi$ is compact while $\tau$ is not.)

We use  ordinary BTZ geometry as the basis for studying the
$\mathbb{Z}_2$ invariant solution in the \emph{covering space}. As
seen in Fig.\ref{O-BTZ} this \z2 invariant geometry is indeed a
double cover of the O-BTZ geometry, i.e. two O-BTZ geometries or two
halves of standard BTZ geometries glued at $r^2=\rho^2=4G\ell^2 M$. %
%which is the ``self-dual'' region under the $\mathbb{Z}_2$
%transformation \eqref{Orientation-Z2}.%
The metric for the double cover of O-BTZ is %
\be\label{Z2-glued-metric}%
\begin{split}%
ds^2 &=[\rho^2 \theta(\Phi)+r^2 \theta(-\Phi)] d\tau^2-8G\ell J
d\tau d\varphi\ \cr &+[r^2\theta(\Phi)+
\rho^2\theta(-\Phi)]d\varphi^2 + \frac{r^2dr^2}{16G^2
J^2-\frac{r^2\rho^2}{\ell^2}}\ ,
\end{split} %
\ee%
where $\theta(X)$ is the step function %,
and%
\be%
\Phi=r^2-4G\ell^2 M=\frac{r^2-\rho^2}{2}. %
\ee%
In the coordinate system \eqref{Z2-glued-metric}  $\tau$ and
$\varphi$ are both dimensionless. The volume element of the geometry
is \be\label{volume-form}\begin{split} dV & =\ell d\tau d\varphi\ (
\theta(\Phi)\ rdr+\theta(-\Phi)\ \rho d\rho )\cr &=\ell r d\tau
d\varphi dr\ (\theta(\Phi)-\theta(-\Phi))\ .
\end{split}\ee%
 That is, as expected and by construction,
the two \ads regions on the opposite sides of the dashed line in
Fig. \ref{O-BTZ} have opposite orientations.

%%%%%%%%%%%%%%%%%%%%%%%%%%%%%%%%%%%%%%%%%%%%%%%%
With the above choice, metric is clearly continuous at $\Phi=0$.
Next, one should make sure that the Israel matching conditions \cite{Israel}
at the junction are also met. For the latter we use the formulation
developed in \cite{Mansouri-Khorrami}. It is straightforward to show
that, in notations of \cite{Mansouri-Khorrami}, the Ricci tensor has the
following jump at $\Phi=0$%
\be\label{Ricci-jump}%
\breve{R}_{\mu\nu}=64G^2(\ell^2M^2-J^2)\ diag(1,0,-1)\ \delta(\Phi)\ ,%
\ee%
in $(\tau,r,\varphi)$ frame. As we see the jump is  caused by a $2D$
object located at $\Phi=0$ with stress tensor $$S_{\mu\nu}=T
\sqrt{g_2}\ diag(1,0,-1),$$ where $g_2=16G^2\ell^2 (\ell^2M^2-J^2)$
is the determinant of the two dimensional $\tau\varphi$ part of
metric \eqref{Z2-glued-metric} at $\Phi=0$, which is the metric at
the junction, and
\be\label{tension}%
T=\frac{1}{4\pi G\ell}\ . %
\ee%
In the region between two horizons $r$ is the time-like coordinate
and hence the orientifold fixed locus is a space-like orientifold
with metric of a cylinder, the SOC and $T$ may be associated with
the tension of the space-like orientifold cylinder. \footnote{The
fact that at the orientifold fixed locus Ricci tensor has a jump is
similar to the $R^2/\mathbb{Z}_k$ orbifold case the analysis of
which has been carried out in \cite{Solodukhin}.}

 Note that the jump
in curvature, and hence the stress tensor \eqref{Ricci-jump},
vanishes for two special cases: massless BTZ ($\ell M=J=0$) and
extremal BTZ ($\ell M=|J|$).  For these two cases the junction is a
light-like cylinder (rather than being space-like) and the
orientifolding does not need an orientifold string of the type
discussed above \cite{progress}.

As the \z2 \eqref{Orientation-Z2} and the BTZ orbifolding
\eqref{t-phi-identification} commute, the O-BTZ solution
\eqref{Z2-glued-metric}  may be considered as a BTZ solution
constructed upon the ``O-AdS$_3$'', the metric for the double cover
of which is
\be\label{new-ads3}%
ds^2 =(-|\tPhi|+\frac{\ell^2}{2}) dt^2%\cr&
-\ell^2 \frac{d\tr^2}{\trho^2}+(|\tPhi|+\frac{\ell^2}{2})d\phi^2 \ ,
\ee%
in coordinate system \eqref{Steif-coordinates}, where
$\tPhi=\tr^2-\ell^2/2$. One may perform the matching condition
analysis in the covering \ads space at $\tPhi=0$ and verify that
they are satisfied upon insertion of a space-like orientifold plane
with tension $T$ given in \eqref{tension} at $\tPhi=0$. The
orientifold fixed plane, however, becomes light-like when we
approach the conformal boundary of AdS$_3$.\footnote{We comment that
\ads in Poincar\'e coordinates,
$ds^2=\ell^2[u^2(-dt^2+dx^2)+\frac{du^2}{u^2}],\ u>0$, which appears
in the near horizon limit of D1-D5 system \cite{ads/cft} can be
extended in a similar way to beyond $u^2>0$ region by replacing
$u^2$ with $|u^2|$. In order to make this a solution to \ads
Einstein gravity we need to add an orientifold cylinder with tension
\eqref{tension} at $u^2=0$. Note that $u=0$ is a light-like (rather
than space-like) direction. }

Remarkably, computing the jump in the value of the density of the
gravity action moving from $II_-$ to $II_+$ region (\emph{cf.}
Fig.\ref{O-BTZ}) one finds
\bea\label{Einstein-action-value}%
\Delta S &=&\frac{1}{16\pi G}\int dV\
(R+\frac{2}{\ell^2})=\frac{1}{\pi} \int d\tau d\varphi
\sqrt{\ell^2M^2-J^2}\cr & =& \int T\  \sqrt{g_2}\ d\tau d\varphi .
\eea

\bc\textit{\textbf{{Discussion and outlook}}}\ec
We have introduced a new class of locally \ads solutions to \ads
Einstein gravity, the O-BTZ black holes. These are BTZ black holes
built upon an \ads with an orientifold projection (O-AdS$_3$) the
fixed locus of  which is a space-like cylinder. Away from the fixed locus the metric for the O-BTZ is exactly the same as ordinary BTZ and hence they
asymptote to  O-AdS$_3$, and are specified by two parameters, ADM
mass $M$ and angular momentum $J$. The orientifold projection used
is chosen such that it keeps the orientation on (conformal) boundary
of AdS$_3$. Having an orientable boundary is important for having a
well-defined 2$D$ CFT with a positive definite energy.

As can be seen from the Penrose diagram depicted in Fig.\ref{O-BTZ},
\footnote{Note that Fig.\ref{O-BTZ} shows Penrose diagram of O-BTZ
geometry in its covering space and that O-BTZ is the region limited
to two successive dashed lines.} O-BTZ geometries do not have the
inner horizon and the region behind it. {}From the viewpoint of an
observer outside the horizon O-BTZ geometry is indistinguishable
from an ordinary BTZ and in particular its horizon area is $2\pi
r_+$, where $r_+$ can be computed in terms of $M$ and $J$ using
\eqref{AdS-MandJ}. Therefore, laws of black hole thermodynamics for
our solutions is written exactly in the same way as for the ordinary
BTZ.

The Brown-Henneaux analysis \cite{Brown-Henneaux} only makes use of
the asymptotic behavior of diffeomorphisms, that the gravity part of
the action is pure \ads Einstein gravity, and is  independent of the
details of the geometry away from the boundary. Therefore, analysis
of \cite{Brown-Henneaux} in a straightforward way generalizes  to
our case and one finds a Virasoro algebra with central charge
\eqref{BH-central-charge} for the O-BTZ solutions \cite{progress}.
This suggests that there is a CFT$_2$ at the central charge
\eqref{BH-central-charge} and that our O-BTZ solutions are states in
this CFT.

Many other aspects of the O-\ads and O-BTZ geometries such as
geodesic motion in these backgrounds,  discussion about the possible
dual CFT$_2$, extension of  proposal made in \cite{Maldacena} for
the initial (Hartle-Hawking) entangled state to the O-BTZ
backgrounds and their relevance for addressing the pure \ads
Einstein gravity problems (\emph{cf.} \cite{PureAdS3-CFT2}) will be
discussed in upcoming publications.

\begin{acknowledgements}
We would like to thank Vijay Balasubramanian, Jan de Boer, Chethan
Krishnan, Kumar S. Narain, Joan Sim\'on, Erik Verlinde for comments and
especially  Kostas Skenderis for  discussions. F.L.
would like to thank IPM for hospitality during the course of this
project. M.M.Sh-J would like to thank the Abdus-Salam ICTP where
this work completed.
\end{acknowledgements}

\vspace*{-5mm}

%**************************************************************************************************************************

\end{document}